\pdfoutput=1
\documentclass[12pt]{article}
\usepackage{epsfig}

\def\beq{\begin{equation}}
\def\eeq#1{\label{#1}\end{equation}}
\def\eeqn{\end{equation}}

\def\beqa{\begin{eqnarray}}
\def\eeqa#1{\label{#1}\end{eqnarray}}
\def\eeqan{\end{eqnarray}}

\let\bar=\overbar

\def\Dslash{\not{\hbox{\kern-4pt $D$}}}
\def\dslash{\not{\hbox{\kern-2pt $\del$}}}

\def\BR{\mbox{\rm BR}}

\def\msb{{\bar{\ssstyle M \kern -1pt S}}}

\input babarsym
\newcommand{\bi}{\begin{itemize}}
\newcommand{\ei}{\end{itemize}}
\newcommand{\ben}{\begin{enumerate}}
\newcommand{\een}{\end{enumerate}}
\newcommand{\bc}{\begin{center}}
\newcommand{\ec}{\end{center}}
\newcommand{\bt}{\begin{table}}
\newcommand{\et}{\end{table}}
\newcommand{\be}{\begin{equation}}
\newcommand{\ba}{\begin{eqnarray}}
\newcommand{\ea}{\end{eqnarray}}

\newcommand{\la}{\ifmmode {\leftarrow} \else {$\leftarrow$}\fi}
\newcommand{\Ra}{\ifmmode {\Rightarrow} \else {$\Rightarrow$}\fi}
\newcommand{\La}{\ifmmode {\Leftarrow} \else {$\Leftarrow$}\fi}
\newcommand{\Lra}{\ifmmode {\Longrightarrow} \else {$\Longrightarrow$}\fi}
\newcommand{\Lla}{\ifmmode {\Longleftarrow} \else {$\Longleftarrow$\fi}}
\newcommand{\Llra}{\ifmmode {\Longleftrightarrow} \else {$\Longleftrightarrow$\fi}}
\newcommand{\Lk}{\ifmmode {{\cal L}} \else {${\cal L}$}\fi}
\newcommand{\Wt}{\ifmmode {{\cal W}} \else {${\cal W}$}\fi}
\newcommand{\Br}{\ifmmode {{\cal B}} \else {${\cal B}$}\fi}
\newcommand{\N}{\ifmmode {{\cal N}} \else {${\cal N}$}\fi}
\newcommand{\G}{\ifmmode {{\cal G}} \else {${\cal G}$}\fi}
\newcommand{\E}{\ifmmode {{\cal E}} \else {${\cal E}$}\fi}
\newcommand{\Pfr}{\ifmmode {{\cal F}} \else {${\cal F}$}\fi}
\newcommand{\Aone}{\ifmmode {{\cal A}_1} \else {${\cal A}_1$}\fi}
\newcommand{\rha}{\ifmmode{\mbox{\rho^2_{{\cal A}_1}}} \else {\mbox{$\rho^2_{{\cal A}_1}$}}\fi}
\newcommand{\rhf}{\ifmmode{\rho^2_{\cal F}}\else{\mbox{$\rho^2_{\cal F}$}}\fi}
\newcommand{\om}{\ifmmode {w} \else {$w$}\fi}
\newcommand{\dom}{\ifmmode {\Delta w} \else {$\Delta w$}\fi}
\newcommand{\tBz}{\ifmmode {\tau_{\Bz}} \else {$\tau_{\Bz}$}\fi}
\newcommand{\tBp}{\ifmmode {\tau_{\Bu}} \else {$\tau_{\Bu}$}\fi}

\newcommand{\psoft}{\ifmmode {\pi_{s}} \else {$\pi_{s}$}\fi}
\newcommand{\plab}{\ifmmode{p} \else {$p$} \fi}
\newcommand{\ctdl}{\ifmmode{ \cos(\theta_{\Dstar\ell}) } \else {$\cos(\theta_{\Dstar\ell})$} \fi}
\newcommand{\ks}{\ifmmode{k^*} \else {$k^*$} \fi}
\newcommand{\mnutag}{\ifmmode{m^2_{\nu ,tag}} \else {$m^2_{\nu ,tag}$}\fi} 
\newcommand{\mnusig}{\ifmmode{m^2_{\nu ,sig}} \else {$m^2_{\nu ,sig}$}\fi} 
\newcommand{\DTau}{\ifmmode {\Delta \tau} \else {$\Delta \tau$}\fi}
\newcommand{\ggcc}{\ifmmode {GeV^2/c^4} \else {$GeV^2/c^4$}\fi}
\def\BpBm {\ensuremath{B^+ {\kern -0.16em \Bub}}}
\def\mm {\ensuremath{{\mathcal{M}_\nu^2}}}
\def\dT{\ensuremath{\Delta t}'}
\def\DT{\ensuremath{\Delta t}}
\def\dZ{\ensuremath{\Delta Z}}

\newcommand{\BtoDs}{\mbox{$\Bzb\rightarrow D^{*+} \ell^- \bar{\nu_\ell}$}\xspace}

\newcommand{\magqp}{\mbox{$|q/p|$}}
\def\poverq2{\ensuremath{\bigg\vert\frac{p}{q}\bigg\vert^2}\xspace}
\def\qoverp2{\ensuremath{\bigg\vert\frac{q}{p}\bigg\vert^2}\xspace}

\def\BzBz     {\ensuremath{\mbox{\Bz {\kern -0.1em \Bz}}}\xspace}
\def\BzBzb     {\ensuremath{\mbox{\Bz {\kern -0.1em \Bzb}}}\xspace}
\def\BzbBzb   {\ensuremath{\mbox{\Bzb {\kern -0.1em \Bzb}}}\xspace}
\def\mm {{\ensuremath{{{\cal M}_\nu}^2}}\xspace}
\newcommand{\Brec}{\mbox{$B_{\mathrm R}$}}
\newcommand{\Btag}{\mbox{$B_{\mathrm T}$}}

\def\Kp{\ensuremath {K^{+}\xspace}}
\def\Km{\ensuremath {K^{-}\xspace}}
\def\Kt{\ensuremath {K_{tag}\xspace}}
\def\Kr{\ensuremath {K_{rec}\xspace}}

\def\At{\ensuremath{\mathcal{A}_{K}}\xspace}
\def\Ar{\ensuremath{\mathcal{A}_{r\ell}}\xspace}

\def\All{\ensuremath{\mathcal{A}_{\CP}}\xspace}

\def\Kt{\ensuremath{K_{T}}\xspace}
\def\Kr{\ensuremath{K_{R}}\xspace}

\def\all{\ensuremath{\mathcal{A}_{\CP}}}

\def\dCP{\ensuremath{\delta_{\CP}}}

\def\dZ{\ensuremath{\Delta Z}}
\def\Zr{\ensuremath{Z_{\mathrm R}}}
\def\Zt{\ensuremath{Z_{\mathrm T}}}
\def\qoverp{\ensuremath{\frac{q}{p}}}

\def\Title#1{\begin{center} {\Large {\bf #1} } \end{center}}

\begin{document}

\Title{$|q/p|$ Measurement from $B^0\rightarrow D^* l \nu$ Partial 
Reconstruction}

\bigskip\bigskip

%+\addtocontents{toc}{{\it D. Reggiano}}
%+\label{ReggianoStart}

\begin{raggedright}  

{\it Martino Margoni\index{Margoni, M.}\\
on behalf of the BaBar Collaboration\\
Universit\`a di Padova and INFN sezione di Padova\\
Padova, ITALY}\\
\bigskip\bigskip

Proceedings of CKM 2012, the 7th International Workshop on the CKM
Unitarity Triangle, University of Cincinnati, USA, 28 September - 2 October 2012
\end{raggedright}

\begin{abstract}
We present
a new measurement of $CP$ violation induced by \Bz\Bzb\
oscillations, based on the full data set collected by the \babar\
experiment at the PEPII collider.
We use a sample of about 6 million \BtoDs\ decays 
selected with partial reconstruction of the \dsp\ meson. The
charged lepton identifies the flavor of the first $B$ meson at its decay time,
the flavor of the other $B$ is determined by kaon tagging.
We determine the parameter 
$\dCP = 1 - |q/p| = (0.29\pm0.84^{+1.78 }_{-1.61})\times 10^{-3}$.
\end{abstract}

\section{Introduction}
\label{sec:intro}
The two-mass eigenstates
of the neutral $B$ meson system, carrying mass $m_L$ and $m_H$, are expressed 
in terms of the flavor eigenstates, \Bz\ and \Bzb, as
$|B_{L,H} \rangle= p |\Bz \rangle \pm q |\Bzb \rangle  $.
%Any deviation from unity of the ratio \magqp\ would imply that the
%mass eigenstates are not $CP$ eigenstates (``mixing-induced $CP$
%violation''). 
%The Standard Model prediction, including NLO QCD corrections, is
%$\dCP = 1 -|q/p|  =  -(2.96 \pm 0.67) \times 10^{-4}$~\cite{Ciuchini}.
%A large deviation from unity would be therefore a clear evidence of New Physics
%beyond the Standard Model.

If $CP$ is violated in mixing, the probability of a \Bz\ to oscillate to
a \Bzb\ is different from the probability of a \Bzb\ to oscillate to a \Bz and
thus we expect to observe a sizeable value for the 
asymmetry:
\begin{equation}\label{eq:A_SL}
%\all = \frac{N(\Bz\Bz) - N(\Bzb\Bzb)}{N(\Bz\Bz) + N(\Bzb\Bzb)} =
%\frac{N(\ell^+\ell^+) - N(\ell^-\ell^-)}{N(\ell^+\ell^+) + N(\ell^-\ell^-)} \smeq
%2 \left( 1 - \left| \frac{q}{p} \right| \right),
%2 \dCP ,
\all = \frac{N(\Bz\Bz) - N(\Bzb\Bzb)}{N(\Bz\Bz) + N(\Bzb\Bzb)} \simeq 2 \dCP .
\end{equation}
where $\dCP = 1 -|q/p|$.

Any deviation from unity of the ratio \magqp\ would imply that the
mass eigenstates are not $CP$ eigenstates (``mixing-induced $CP$
violation''). 
The Standard Model prediction is
$\all =  -(4.1 \pm 0.6) \times 10^{-4}$~\cite{Lenz}.
A large deviation from $\dCP=1$ would be therefore a clear evidence of New Physics
beyond the Standard Model.

%$B$-factories \cite{BaBar_dilep}, \cite{Belle_dilep}  and hadron
%collider experiments have published results
%where the $B$ mesons are tagged by their semileptonic decays.
%The $D\emptyset$ Collaboration observes a large deviation from the SM,
%which is however attributed to \Bs\ meson oscillation \cite{D0_mumu}.
%LHCb preliminary result is in fact compatible both with
%the SM and with $D\emptyset$  \cite{LHCb} .

%These dilepton measurements benefit from the large amount of events which
%can be selected at $B$-factories or at hadron colliders. They however rely on
%the use of control samples to subtract the charge asymmetric
%background from hadron to lepton misidentification or light hadron
%decay, and to compute the charge dependent lepton identification
%asymmetry which may produce a fake signal. These systematic
%uncertanties constitute a severe limitation to the precision of the measurement.

%We present here a new kind of measurement.

\section{Analysis Method}
\label{sec:method}
We present a measurement based on the partial reconstruction of \BtoDs\ 
decays (hereafter \Brec). 
%where we identify only the
%charged lepton and the low momentum pion (\psoft) from the 
%$\dsp \rightarrow \Dz \psoft$ decays. 
A state decaying as a \Bz (\Bzb)
meson produces a positive (negative) charge lepton. 
The observed asymmetry between the number 
of positive-charge and negative-charge leptons is therefore:
\begin{eqnarray}
\label{eq:arec}   A_\ell \simeq \Ar + \All  \chi_d,
\end{eqnarray}
where $\chi_d=0.1862\pm0.0023$ \cite{PDG} is the integrated mixing 
probability for \Bz\
mesons, and \Ar\ is the charge asymmetry in the reconstruction of
\BtoDs\ decays.

We use kaons from decays of the other \Bz\ (\Btag) to tag its
flavor (\Kt).
 A state decaying as a \Bz (\Bzb) meson results most often in a \Kp (\Km). If
mixing takes place, the $\ell$ and the \kaon\ have then the same electric
charge. The observed asymmetry in the rate of mixed events is:
\begin{eqnarray}
\label{eq:amix}   A_{T} = \frac{N(\ell^+\Kt^+) - N(\ell^-\Kt^-)} {N(\ell^+\Kt^+) + N(\ell^-\Kt^-)} \simeq \Ar + \At + \All ,
\end{eqnarray}
where \At\ is the charge asymmetry in kaon reconstruction.
A kaon with the same charge as the $\ell$
might also come from the Cabibbo-Favored (CF) decays  of the \Dz\
meson produced with the lepton from the partially reconstructed side (\Kr).
The asymmetry observed for these events is:
\begin{eqnarray}
\label{eq:asame} A_{R} =\frac{N(\ell^+\Kr^+) - N(\ell^-\Kr^-)} {N(\ell^+\Kr^+) + N(\ell^-\Kr^-)} \simeq \Ar+\At + \All \chi_d
\end{eqnarray}

Eqs. \ref{eq:arec}, \ref{eq:amix}, and \ref{eq:asame} can be
inverted to extract \All\ and the detector induced asymmetries.
%It is not possible to distinguish in each event a \Kt from a 
%\Kr. They are separated on statistical basis, using
%kinematics features and proper-time difference information.
%We perform a multidimensional binned-likelihood fit to
%determine, together with the asymmetries, several other factors 
%which would be otherwise sources of systematic uncertainty.

\section{Extraction of $\dCP$}
\label{sec:method}

The data sample used in this analysis consists of an integrated luminosity of 425.7\invfb, 
corresponding to 468 million \BB pairs, collected at the $\Y4S$
resonance by the \babar\ detector. 
%A detailed description of the \babar\ detector and the algorithms used
%for charged and neutral particle reconstruction and identification is provided
%elsewhere~\cite{babar_nim}.
%A sample of 
%45\invfb collected 40\mev below the resonance are used to describe the non-\BB\ (continuum) ba%ckground.
%A simulated sample of $\BB$ events with integrated luminosity equivalent to approximately 
%three times the size of the data sample is also used.

%We preselect a sample of hadronic events with at least four charged tracks.
%To reduce continuum background, we require that the ratio of the 2$^{nd}$ to the
%0$^{th}$ order Fox-Wolfram~\cite{wolfram} variables be less than 0.6.
We select a sample of partially reconstructed $B$ mesons 
%in the channel 
%$\Bzb \rightarrow D^{*+} X \ell^{-} \bar{\nu}_{\ell}$, 
by retaining events containing a charged lepton ($\ell = e,\,\mu$) and a low momentum 
pion (soft pion, $\pi^+_{s}$) from the decay $D^{*+}\to \Dz \pi^+_{s}$.
The lepton momentum must be in the range $1.4 < p_{\ellm} < 2.3 \gevc$ and 
the soft pion candidate must satisfy $60 < p_{\pi^{+}_{s}} < 190 \mevc$.
Throughout the paper the momentum, energy and direction of all particles are
computed in the $e^+e^-$ rest frame.
%The two tracks must be consistent with originating from a common vertex, constrained to the 
%beam-spot in the plane transverse to the beam axis. Finally, we
%combine $p_{\ellm}$, $p_{\pi^{+}_{s}}$ 
%and the probability from the vertex fit into a likelihood ratio
%variable ($\eta$), 
%optimized to reject \BB\ background. If more than a combination is
%found in an event, we keep 
%that 
%the one
%with the largest value of $\eta$.

Using conservation of momentum and energy, 
the invariant mass squared of the undetected neutrino is calculated as
%
%\begin{eqnarray}
$\mm \equiv (E_{\mbox{\rm \small beam}}-E_{{D^*}} - 
E_{\ell})^2-({\vec{p}}_{{D^*}} + {\vec{p}}_{\ell})^2 ,
$
%\label{eqn:mms}
%\end{eqnarray}
%
where $E_{\mbox{\rm \small beam}}$ is half the total center-of-mass energy and $E_{\ell}~(E_{{D^*}})$ 
and ${\vec{p}}_{\ell}~({\vec{p}}_{{D^*}})$ are the energy and momentum 
of the lepton (the $D^*$ meson).  
%Since the magnitude of the $B$ meson
%momentum, $p_{B}$, is sufficiently 
%small compared to $p_{\ell}$ and $p_{D^*}$, we set
%$p_{B}$ = 0.
% in obtaining Eq.~\ref{eqn:mms}.
%As a consequence of the limited phase space available in the $D^{*+}$
%decay, the soft pion is emitted nearly at rest in the $D^{*+}$ rest frame.
The $D^{*+}$ four-momentum can be computed by approximating 
its direction as that of the soft pion, and parameterizing its momentum as 
a linear function of the soft-pion momentum.
We select pairs of tracks with opposite electric charge for our signal ($\ell^\mp \psoft^\pm$)
and we use same-charge pairs ($\ell^\pm \psoft^\pm$) for background studies.  

%Several processes where $\dsp$ and $\ell^-$ originate from the same $B$-meson produce 
%a peak near zero in the \mm\ distribution. The peaking signal consists
%of  (a) $\Bzb \rightarrow D^{*+} \ell^{-} \bar\nu_{\ell}$ decays (primary);
%(b) $\Bzb \rightarrow D^{*+} (\mathrm{n}\pi) \ell^- \bar{\nu}_{\ell}$
%(\dstrstr) , (c) $\Bzb \rightarrow D^{*+}\tau^- \bar{\nu}_{\tau} $, $\tau^- \rightarrow
%\ell^{-}\bar{\nu}_{\ell}\nu_{\tau} $. The main source of peaking
%background is due to charged-B decays to excited  resonant or non
%resonant charm excitations, $\Bp \rightarrow D^{*+}
%(\mathrm{n}\pi) \ell^- \bar{\nu}_{\ell}$, or to $\tau$ leptons, fake lepton 
% $B \rightarrow D^{*+} h^- X$, where the hadron ($h =
% \pi,K, D)$ is erroneously identified as, or decays to, a
% charged lepton (fake-lepton).
% We also include radiative events, where photons with energy
%above 1 MeV are emitted by any charged particle, as described by PHOTOS~\cite{photos} in our s%imulation. 
%We define the signal region $\mm > -2~$GeV$^{2}/c^4$, and the sideband $-10 < \mm < -4~$GeV$^{%2}/c^4$.

\begin{figure}[htb]
\begin{center}
\epsfig{file=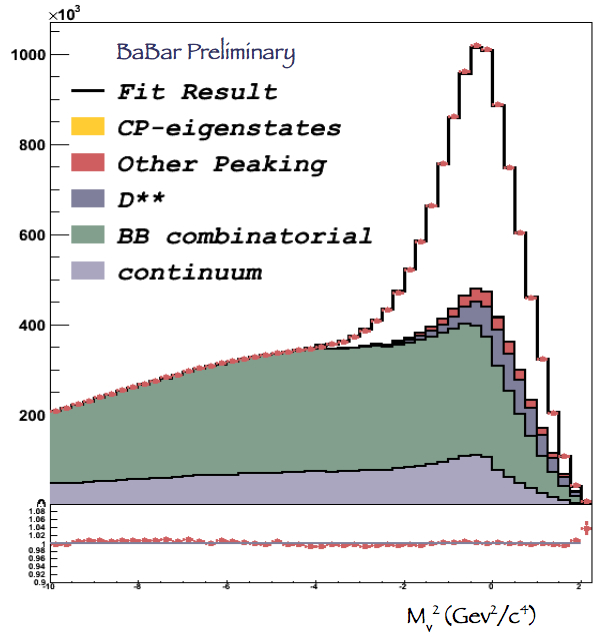,height=3.1in}
\caption{\mm\ distribution for the data, points with error
  bars, and the fitted contributions from the various sample components.}
%from signal, peaking background,
%  \BB\ combinatorial and continuum.}
\label{f:fitmm}
\end{center}
\end{figure}

%Light quark (continuum) events and random combinations of a low
%momentum pion and an opposite charge lepton from
%combinatorial \BB\ events, contribute to the non-peaking background.
We determine the number of signal events in our sample with a minimum
$\chi^2$ fit to the \mm\ distribution in the 
interval $-10 <\mm< 2.5~$GeV$^2$/c$^4$. 
%In the fit, the continuum contribution is obtained from off-peak
%events, normalized by the on-peak to off-peak luminosity ratio, the
%other contributions are taken from the simulation. 
%The amount of
%events from combinatorial \BB\ background, primary decays and \dstrstr\ are allowed to vary in% the
%fit, while the other peaking contributions (few percent) are fixed to the simulation
%expectations, rescaled by the luminosities ratios. The amount of \Bz\
%mesons in the sample is then obtained assuming that 2/3 of the fitted
%amount of \dstrstr\ events are produced by \Bp\ decays, as suggested by simple
%isospin considerations. 
A total of  $(5945\pm 7) \cdot 10^3$ peaking events are found.
% in the
%full range peaking events account for about 30\%\ of the sample,
%continuum background for about 15\%.
The result of the fit is displayed in
Fig.\ref{f:fitmm}.

%\vspace*{-0.5cm}
%\begin{center} 
%\begin{figure}[htbp]
%\includegraphics[width=8cm]{FitResult.eps}  
%\vspace*{-1.5cm}
%\caption{\mm\ distribution for the data, points with error
%  bars, and the fitted contributions from signal, peaking background,
%  \BB\ combinatorial and rescaled off-peak events (continuous line
%  overlied).}
%\label{f:fitmm}
%\end{figure}
%\end{center}
%\vspace*{-1cm}

%We select kaons from all the charged tracks with momentum
%larger than 0.2 \gevc\  combining informations from the Cherenkov detector with the
%measurements of the energy losses in the trackers. True kaons are
%identified with 86\% efficiency and 3.4\% pion mis-identification
%rate. 
%Kaons may be produced from the decay of the \Dz\ from the
%partially reconstructed \Bz\ (\Kr), or in any step of the decay of the
%other B (\Kt).
%We exploit the relation between the charge of the lepton and that of
%the \Kt\ to tag mixing.
%When an oscillation takes place, a \Kt\ from a Cabibbo Favored (CF) decay has the
%same charge as the $\ell$. 
%Equal-charge combinations are also
%observed from Cabibbo Supressed (CS) \Kt production in unmixed events, and from CF \Kr\ produc%tion.
%Unmixed CF \Kt, mixed CS \Kt, and CS \Kr, result in opposite-charge
%combinations. 
%Fake kaons contribute both to equal and opposite charge events with 
%comparable rates. 

We define $\dZ = \Zr - \Zt$,
where \Zr\ is the projection along the beam direction of the \Brec\
decay point, and \Zt\ is the projection along the same direction of the 
intersection of the \kaon\ track trajectory with the beam-spot.
%In the boost approximation \cite{BAPX} 
We measure the
proper-time-difference between the two $B$ mesons using the relation
$\DT = \dZ /( \beta \gamma c)$, where the parameters $\beta,\gamma$ 
express the Lorentz Boost from the laboratory to the \FourS\ rest frame.
%, are determined run% by run from PEPII settings.
%We reject events if the error $\sigma(\DT)$ exceeds 3 ps.

We distinguish \Kt\  from \Kr\ using proper-time difference and kinematic informations.
Due to the short lifetime and small boost of the \Dz\ meson, 
small values of \DT\ are expected for the \Kr. 
%Much larger values are instead expected for CF mixed \Kt,
%due to the long period of the \Bz\ oscillation. 
%(about six times the
%\Bz\ lifetime).  
%By fitting the \DT\ distribution for equal and
%opposite charge $\ell$-K combinations, we also compute the
%contamination from CS \Kt\ decays.
%To improve the separation between \Kt\ and \Kr, we also exploit kinematics. 
The $\ell$ and the \dsp\ are emitted
at large angles in the rest frame of the decaying \Bz : therefore the angle $\theta_{\ell\kaon}$
between the $\ell$ and the \Kr\ has values close to $\pi$, and
cos$(\theta_{\ell\kaon})$ close to $-1$. The  corresponding distribution
for the \Kt\ is uniform.

%If more than a Kaon is found in an event, we consider each
%combination in turn. We use parameterized simulations (toys) to verify the
%effect of this choice on the computation of the statistical uncertainty.

The measurement proceeds in two steps. 
We first measure the sample
composition of the eight
tagged samples defined by lepton type, lepton charge and $K$
charge, with the fit to \mm\ described above. 
%We also fit the four
%inclusive lepton samples to determine the charge asymmetries at the
%reconstruction stage (see eq. \ref{eq:arec}).
The results of the first stage are used in the second
stage, 
where we perform
a binned maximum likelihood fit to a
two-dimensional PDF obtained as a  product
of the \DT\ and cos($\theta_{\ell K})$ functions.
 
%fit simultaneously the cos$\theta_{\ell K}$ and \DT\
%distributions in the eight tagged samples. 
%The result is obtained with 
%a binned maximum likelihood fit to a
%two-dimensional PDF obtained as a  product
%of the \DT\ and cos($\theta_{\ell K})$ functions. 

%Off-peak events are interpolated to parameterize the
%continuum distribution. 
The \DT\ distributions for \Kt\ \BB\ events are
parameterized as the convolutions of the theoretical distributions
${\cal F}_i(\dT|\vec{\Theta})$ with
the resolution function $ {\cal R} (\DT,\dT) $: ${\cal G}_i(\DT) = \int_{-\infty}^{+\infty}
{\cal F}_i(\dT|\vec{\Theta}) {\cal R} (\DT,\dT) d(\dT)$, where $\dT$
is the actual difference between the times of decay of the two mesons and
$\vec{\Theta}$ is the vector of the physical parameters.

The resolution function 
%accounts for the
%experimental uncertainties in the measurement of \DT, for the smearing
%due to the boost approximation, and for the displacement of the \Kt\
%production point from the \Btag\ decay position due to the motion of the
%charm meson. 
%It 
consists of the superposition of several Gaussian
functions convolved with exponentials. We use a different set of
parameters for peaking and for combinatoric events. 

To describe the \DT\ distributions for \Kr\ events, ${\cal G}_{\Kr} (\DT)$,  we select a
sub-sample of data containing less than 5\% \Kt\ decays, and we use the
background subtracted histograms in our likelihood. 
%As an alternative, we apply the same selection to the simulation and we
%correct the \DT\ distribution predicted by the Monte Carlo by the
%ratio of the histograms extracted from data and simulated events.
%Simulation shows that the distributions so obtained are unbiased.
%We take the average of the two $\dCP$ determinations obtained with the two 
%different strategies as our nominal result.

%Continuum events (${\cal G}_{cnt} (\DT)$) are represented by a decaying 
%exponential, convoluted
%with a resolution function similar to that used for $B$ events. 
%The
%effective lifetime and resolution parameters are determined 
%by fitting simultaneously the
%off-peak data.

%The result is obtained with a binned maximum likelihood fit to a
%two-dimensional PDF obtained as a  product
%of the \DT\ and cos($\theta_{\ell K})$ functions. 

The individual
cos($\theta_{\ell K}$) shapes are obtained 
from the histograms of the
simulated distributions for \BB\ events, separately for \Kt\ and \Kr\
events. 

Events belonging to each of the eight tagged samples are
grouped in 100 \DT\ bins, 25 $\sigma(\DT)$ bins, 4 cos$\theta_{\ell,K}$
bins, and 5 \mm\ bins. We further split data in five bins of $K$
momentum, $p_K$, to account for the dependencies of several parameters,
describing the \DT\ resolution
function, the cos$(\theta_{\ell K})$ distributions, and the fractions
of \Kt\ events, observed in the simulation. 

The rate of events in each bin ($\vec{j}$) and for each tagged sample are then
expressed as the sum of the predicted contributions from peaking
events, \BB\ combinatorial and continuum background.
%{\small
%\begin{eqnarray} 
%{\cal N}_{\ell\kaon} (\vec{j}) &=& {\cal N}[
 % (1-f_{\Bp}-f_{$CP$e}-f_{cmb}-f_{cnt}) {\cal G}_{\Bz}(\vec{j}) \\
%\nonumber
%&+& f_{\Bp} {\cal G}_{\Bp}(\vec{j})  
% + f_{CPe} {\cal G}_{CPe}(\vec{j}) \\
%\nonumber
% &+& f^0_{cmb} {\cal G}_{\Bz,cmb}(\vec{j})
% + f^+_{cmb} {\cal G}_{\Bp,cmb}(\vec{j})
% + f_{cnt} {\cal G}_{cont}(\vec{j}) ]
%\end{eqnarray}
%}
%where the fractions of peaking, $CP$ eigenstates, combinatoric \BB\ , and continuum
%events in each \mm\ interval is fixed to the results of the first
%stage. The amounts of \Bz\ and of \Bp\ events in the combinatoric background are assumed from
%the simulation.

Accounting for mistags and \Kr\ events, the peaking \Bz\
contributions 
to the equal-charge samples are:
%\begin{widetext}
\begin{eqnarray}
\nonumber 
%{\small
{\cal G}_{\ell^\pm K^\pm} (\vec{j}) =  (1\pm\Ar)(1\pm\At)  
\{(1-f_{\Kr}^{\pm\pm}) [(1-\omega^\pm) {\cal
  G}_{\Bz\Bz / \Bzb \Bzb } (\vec{j}) +
\\
\omega^\mp
 {\cal G}_{\Bz  \Bzb/ \Bzb \Bz}(\vec{j}) ] + f_{\Kr}^{\pm\pm}
 (1-\omega'^{\pm}){\cal G}_{\Kr}(\vec{j}) (1\pm\chi_d \All)~\} 
%}
\end{eqnarray}
%\end{widetext}
where the reconstruction asymmetries are computed separately for the
$e$ and $\mu$ samples.  We allow for different mistag probabilities
for \Kt\ ($\omega^\pm$) and \Kr\ ($\omega'^\pm$).
%because the former come from a mixture of $D$ mesons, while the others
%are produced by \Dz\ decays only.

The $f_{\Kr}^{\pm \pm}(p_k)$ parameters describe the
fractions of \Kr\ tags in each sample in terms of 
the kaon momentum.  
%We let the fit
%determine the values of the $f_{\Kr}^{\pm \pm}$ parameters in each
%$p_k$ bin.
A total of 171 parameters are determined in the fit.

After unblinding we find:
$\dCP = 1 - |q/p| = (0.29\pm0.84^{+1.78}_{-1.61})\times 10^{-3}$.

We consider several sources of systematic errors: 
uncertainties on the composition of the peaking and the combinatorial samples, 
on the description of the \Kr\ and the $CP$ eigenstate \DT\ distributions, and on the
$\Delta\Gamma$ and $\Delta m_d$ parameters. 
Parameterized simulations are used to check the estimate
of the result and its statistical uncertainty. The difference between 
the nominal 
result and the average of those obtained from pseudo-experiments 
is included in the total systematic uncertainty as well as the
statistical error of a validation test performed using the simulation.

%We determine the systematic error due to a possible fit bias by adding the
%statistical error of a validation test performed using the simulation 
%with the difference between the nominal result and the central value
%of the pseudo-experiments ones.

%The value of \dmd ($=0.5085\pm0.0009$) is consistent with the world
%average, while
%the value of \tBz ($=1554\pm0.002$)  is slightly larger than
%expected, an effect also observed in the simulation. 
%By fixing its value to the world average, the $|q/p|$ result increases by
%$0.18\times 10^{-3}$. This effect is taken into account in the 
%systematic error computation. 
%A sizable asymmetry is observed at the reconstruction stage, for both
%$e$ ($30\pm4\times10^{-4}$) and $\mu$ ($31\pm5\times10^{-4}$) , and at
%the \kaon\ tagging stage ($1.37\pm0.03\%$).

\begin{center} 
\begin{figure}[htbp]
\vspace*{-0.5cm}
\includegraphics[width=8cm]{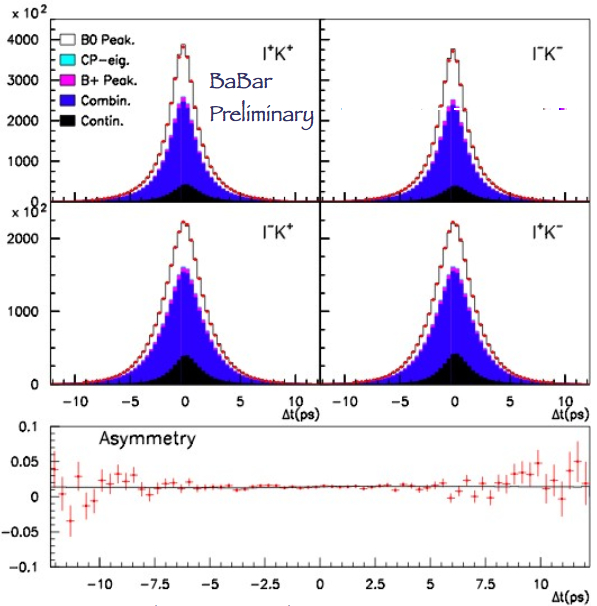}  
\includegraphics[width=8cm]{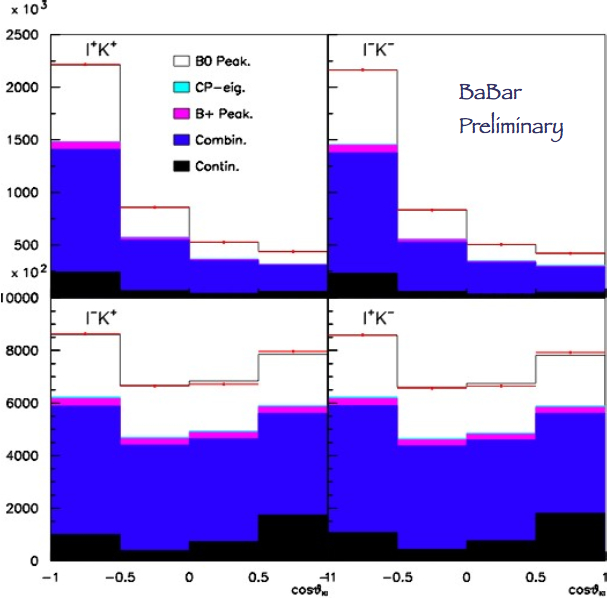}  
\vspace*{-0.5cm}
\caption{\DT\ (left) and cos$\theta_{\ell K}$ (right) distributions for the data, 
point with error bars, and the fitted 
contributions from the various sample components.
%signal, peaking $B^+$ background, $CP$-eigenstates, \BB\ 
%combinatorial and continuum events.
%Top left plot: $\ell^+ K^+$.
%Top right plot: $\ell^- K^-$. 
%Central left plot: $\ell^- K^+$ events. 
%Central right plot: $\ell^+ K^-$ events . 
Bottom left plot: Raw asymmetry between 
$\ell^+ K^+$ and $\ell^- K^-$ events.}
\label{f:Dz}
\end{figure}
\end{center}

Fig. \ref{f:Dz} shows the fit projections for \DT\
and cos$\theta_{\ell K}$.  

%\begin{center} 
%\begin{figure}[htbp]
%\includegraphics[width=8cm]{costhe.jpg}  
%\caption{cos$\theta_{\ell K}$ distribution for the data, point with error bars 
%and the fitted 
%contributions from the various sample components.
%signal, $CP$-eigenstates, peaking $B^+$ background, \BB\ 
%combinatorial and continuum events.
%Top left plot: $\ell^+ K^+$.
%Top right plot: $\ell^- K^-$. 
%Central left plot: $\ell^- K^+$ events. 
%Central right plot: $\ell^+ K^-$ events.} 
%\label{f:costhe}
%\end{figure}
%\end{center}
In summary, we present a new precise measurement of the parameters governing
$CP$ violation in \Bz \Bzb oscillations. With a 
partial \BtoDs\ reconstruction and \kaon\ tagging we find 
$\dCP = (0.29\pm0.84^{+1.78 }_{-1.61})\times 10^{-3},$ and
$ \All = (0.06\pm0.17^{+0.36}_{-0.32})\%.$
These results are consistent with and more 
precise than the \B-factories results from dilepton measurements. 
No deviation is observed from the SM expectation~\cite{Lenz}.

%\include{conc}

%%%%%%%%%%%%%%%%%%%%%%%%%%%%%%%%%%%%%%%%%%%%%%%%%%%%%%%%%%%%%%%%%%%%%%%%%
%%
%%   use this format to include an .eps figure into your paper
%%
%\begin{figure}[htb]
%\begin{center}
%\epsfig{file=magnet.eps,height=1.5in}
%\caption{}
%\label{fig:magnet}
%\end{center}
%\end{figure}
%%%%%%%%%%%%%%%%%%%%%%%%%%%%%%%%%%%%%%%%%%%%%%%%%%%%%%%%%%%%%%%%%%%%%%%%%%%

%%%%%%%%%%%%%%%%%%%%%%%%%%%%%%%%%%%%%%%%%%%%%%%%%%%%%%%%%%%%%%%%%%%%%%%%%
%%
%%   use this format to include a LaTeX table  into your paper
%%
%\begin{table}[b]
%\begin{center}
%\begin{tabular}{l|ccc}  
%Patient &  Initial level($\mu$g/cc) &  w. Magnet &  
%w. Magnet and Sound \\ \hline
% Guglielmo B.  &   0.12     &     0.10      &     0.001  \\
% Ferrando di N. &  0.15     &     0.11      &  $< 0.0005$ \\ \hline
%\end{tabular}
%\caption{Blood cyanide levels for the two patients.}
%\label{tab:blood}
%\end{center}
%\end{table}
%%%%%%%%%%%%%%%%%%%%%%%%%%%%%%%%%%%%%%%%%%%%%%%%%%%%%%%%%%%%%%%%%%%%%%%%%%%


\bigskip

\begin{thebibliography}{99}

%%
%%  bibliographic items can be constructed using the LaTeX format in SPIRES:
%%    see    http://www.slac.stanford.edu/spires/hep/latex.html
%%  SPIRES will also supply the CITATION line information; please include it.
%%

%\bibitem{Mesmer}
%F. A. Mesmer, Proc. Wien. Acad. Sci. {\bf 13}, 1564, 1593 (1762).
%%CITATION = PWASA,13,1564;%%

\bibitem{Lenz} A. Lenz, arXiv:1102.4274.
%\bibitem{Ciuchini} M.~Ciuchini {\em et al.}, 
%J. High Energy Phys. 0308, 031 (2003).
%\bibitem{BaBar_dilep} B. Aubert {\em et al.} (\babar\ collaboration), 
%Phys. Rev. Lett. 96, 251802 (2006). 
%\bibitem{Belle_dilep} E. Nakano {\em et al.} (Belle collaboration), 
%Phys. Rev. D73, 112002 (2006).   
%\bibitem{D0_mumu} V. M. Abazov {\em et al.} (D0 collaboration), 
%Phys. Rev. D84, 052007 (2011).
%\bibitem{LHCb} The LHCb collaboration, LHCb-CONF-2012-022 (2012).
\bibitem{PDG} J. Beringer {\em et al.}, (Particle Data Group), 
Phys. Rev. D86, 010001 (2012).
%\bibitem{babar_nim} B. Aubert {\em et al.} (\babar\ collaboration),
%Nucl. Instr. and Meth. in Phys. Res. A479, 1 (2002).
%\bibitem{wolfram} G. C. Fox and S. Wolfram, Phys. Rev. Lett. 41, 1581 (1978).
%\bibitem{frame} Throughout the paper the momentum, energy and direction of 
%all particles are computed in the $e^+e^-$ rest frame.
%\bibitem{photos} E. Barberio, B. van Eijk, and Z. Was, Comput. Phys. Commun.
%66, 115 (1991); E. Barberio and Z. Was, Comput. Phys. Commun. 79, 291 (1994).
%\bibitem{BAPX} The \babar\ Physics Book, SLAC Report SLAC-R-504.
%\bibitem{DCS} O. Long {\em et al.} Phys. Rev. D68, 034010 (2005).
%\bibitem{UT} A. A. Logunov {\em et al.}, Phys. Lett. B 24, 181 (1967);
%K. G. Chetyrkin and N. V. Krasnikov, Nucl. Phys. B119, 174 (1977);
%K. G. Chetyrkin {\em et al.}, Phys. Lett. B 76, 83 (1978).
%\bibitem{Lenz} A. Lenz, arXiv:1205.1444.

\end{thebibliography}
\end{document}